\documentstyle[aps]{revtex}

\begin{document}

\title{Synchronization and multi-mode dynamics of mutually coupled
       semiconductor lasers}

\author{Claudio R. Mirasso$^{1,2}$,
        Miroslav Kolesik$^{1,3}$,
        Marcelo Matus$^{1}$,
        J.K. White$^{4}$,
        and Jerome V. Moloney$^{1}$ }

\address{$^{1}$ Arizona Center for Mathematical Science,
         University of Arizona, Tucson, AZ 85721.}

\address{$^{2}$ Departament de F\'{\i}sica, Universitat de les Illes Balears,
         E-07071 Palma de Mallorca, Spain}

\address{$^{3}$ Institute of Physics, Slovak Academy of Sciences, Bratislava, Slovakia.}

\address{$^{4}$ Nortel Networks, High Performance Optical Solutions, 3500 Carling
         Ave. Ottawa ON K2H 8E9, Canada.}

\maketitle

\begin{abstract}

{\bf Abstract: } Dynamics of coupled semiconductor lasers is
investigated by numerical simulations. A realistic laser
simulation engine is used to study the synchronization and
dynamical regime in two mutually coupled Fabry-Perot and/or DFB
lasers. Both, single- and multi-mode operation regimes are studied
with emphasis on the role of the multiple laser-cavity modes. Our
findings indicate that the two laser synchronize within each
laser-cavity mode, while the synchronization across different
cavity modes is significantly weaker.
\end{abstract}

\bigskip

Synchronization between coupled nonlinear oscillators has recently
attracted attention of many researchers. A rich palette of
behaviors has been observed in a wide variety of systems
including, among others, population dynamics, coupled neurons, and
lasers\cite{Strogatz93,Ernst95,Roy94,tilmann,White99}. The
interest in the synchronization between chaotic semiconductor
lasers has been motivated by its potential for practical
applications for example in communication systems using chaos to
camouflage the transmitted messages\cite{raj98,Fischer00}. In most
of the cases the coupling between the subsystems includes a delay
that accounts for the time the information takes to travel from on
subsystem to the other. This delay introduces additional degrees
of freedom to the system and leads to a qualitative different
dynamics. The effect of the delay between two mutually coupled
semiconductor lasers has been studied recently\cite{tilmann,josep}
in a regime of long delays and moderate injection couplings. A
spontaneous symmetry breaking was observed together with a
retarded synchronization of chaotic regimes between the two
subsystems. Similar studies, but with weak couplings and short
delay times, have demonstrated localized synchronization of
relaxation oscillations \cite{angela}.

A semiconductor laser model, described by partial differential
equations and including  parabolic gain model, was used for the
first time in ref. \cite{White99} to study feedback effects. In
the latter paper, it was shown that weak external feedback can
promote multi-longitudinal mode instabilities in an otherwise
nominally single mode semiconductor laser, a fact that was also
observed experimentally \cite{Vaschenko}. Moreover, in ref.
\cite{White99} it is found that when two identical semiconductor
lasers subjected to optical feedback are coupled unidirectionally,
the same individual laser cavity mode could synchronize to its
counterpart eventhough the others modes might be out of
synchronization. Despite the relevance that multimode behaviour
may have in some cases, majority of numerical simulations have
been carried out with the rate equations model that assumes a
single laser-cavity mode operation and neglect spatial
dependencies. Although the agreement between the rate-equation
based models and experimental observations is very good in
general, questions concerning the role of the multi-mode laser
operation arise\cite{Vaschenko}. Though it is possible to extend
the Lang-Kobayashi-type models to multi-mode
systems\cite{Viktorov}, we feel that the underlying approximations
are difficult to control and prefer a more direct approach. The
present work aims at further understanding of the multi-mode
behavior in mutually coupled lasers by numerical simulations that
are free of the usual rate-equation model approximations. To this
end, we use a laser simulator with a full spatial and temporal
resolution\cite{simulator}. The simulation engine allows us to
perform realistic numerical experiments on systems consisting of
various types of semiconductor lasers as well as passive cavities
and coupling between the subsystems.

We consider two identical devices, which will be either pairs of
Fabry-Perot or DFB lasers. We pump both lasers with the same
injection current very close to their solitary threshold. The
distance between the two lasers is set to 1.2 m, or equivalently a
flight time for the light of $\tau \sim 4$ ns. A neutral density
filter reduces the coupling between the lasers, which we fix to a
value of 6\% of transmission. For the Fabry-Perot laser we
consider devices of 250 $\mu$m length and 4 $\mu$m width with
natural, as-cleaved, facet reflectivities. In the case of DFB
lasers, we use slightly longer devices (400 $\mu$m) with simple
Bragg gratings, i.e. without grating phase inserts, with a
coupling coefficient of 5000 m$^{-1}$ and zero reflectivity at the
facets. Though such devices exhibit two symmetric grating
supported modes, due to the asymmetry induced by the coupling with
the counterpart laser, one of the modes is greatly suppressed. We
choose the parameters of both types of lasers such that they
operate with carrier densities at which the active layer exhibits
the alpha factor around 3.

The simulational model includes full many-body microscopic gain
and refractive index and correctly accounts for gain dispersion in
a broad frequency band. The flexibility of the simulator allows us
to consider both Fabry-Perot type and DFB type of lasers without
any restrictions of their modal properties. The simulator also
accounts for both mutual injection and feedback coming from the
front facet of the counterparting laser. Moreover, the simulator
also allows us to check situations in which the optical feedback
has negligible effects, as reported in \cite{tilmann},
\cite{angela}, but we can anticipate that the results do not
qualitatively change. Under these conditions we approach as much
as possible the situation reported in recent experiments
\cite{tilmann}, \cite{angela}.

The main features we observe in the simulations with the
Fabry-Perot laser twins can be summarized as follow:

\begin{enumerate}
\item{} We observe a LFF behavior, characterized by a sudden drop of the
total intensity, similar to the one reported in refs.
\cite{tilmann},\cite{josep}. This behavior resembles the well
known LFF regime that appears in the case of a laser subjected to
optical feedback (see for example \cite{guido} and references
therein). However, we have observed that this regime persists even
when we exclude feedback effects from the facet of the other
laser, which is an indication that mutual injection alone may
induce this kind of instabilities.

\item{} We observe a well defined leader-laggard dynamics, as
reported experimentally and numerically with a rate equation model
\cite{tilmann},\cite{josep}, where the role of the leader and
laggard changes randomly from one dropout to the other.

\item{} We observe a high degree of synchronization between the total
output power of both lasers when one of the outputs is shifted
with respect to the other by a time $\tau$, $\tau$ being the time
it takes the light to fly from one laser to the other.

\item{} We only observe significant degree of synchronization if one of
the series is shifted with respect to the other by an integer, but
odd, multiple of $\tau$.
\end{enumerate}

In Figure~1 we show the typical time traces of the total output
power and the optical spectra of both lasers, the latter being
averaged over a whole LFF cycle. In panel a) the output power of
both laser exhibit the LFF features that we have already
mentioned. As expected, fast irregular pulsations, in the GHz
range, develop within these slow LFF cycle. In panel b) it can be
seen that the lasers operate in a multi-mode regime. Despite this
complicated dynamics, the spectra of both lasers are so similar to
each other that it is very difficult to distinguish them. This is
an indication of synchronization between the two lasers. However,
these spectra do not tell us much about the dynamical evolution of
the individual longitudinal modes. To gain insight into this
problem we concentrate on the dynamics that take place within the
different longitudinal modes of the FP lasers. To resolve the
modes, we use a Fabry-Perot filter with a FHWHM  bandwidth of 10
GHz that allows us to isolate each individual longitudinal
laser-cavity mode. In Figure~2 we plot the temporal evolution of
one of the main mode's power for both lasers for a time interval
that corresponds to the range $\sim$ 100 - 200 ns of panel a) of
Figure~1. When one of the series is shifted by $\tau$ a
well-synchronized dynamics can be observed.

To characterize quantitatively the degree of synchronization
between the different longitudinal modes of the two lasers, we
compute the cross correlation function between the same
longitudinal mode of the two lasers, defined as:
\begin{eqnarray*}
S_i(\Delta t) & = & \frac{<\delta P_1^i(t) \delta P_2^i(t-\Delta
t)>} {\sqrt{<(\delta P_1^i(t))^2> <(\delta P_2^i(t))^2>}}
\end{eqnarray*}
where $P_1^i(t)$ and $P_2^i(t)$ represent the output power of the
i-th longitudinal mode of each laser. Figure~3 shows the cross
correlation function between the total power, the power of one of
the main modes (located at $\sim$ -200 GHz in figure 1 b)) and the
power of one side mode (in this case the one located at $\sim$
-600 GHz in figure 1 b)). In all the cases we observe maxima of
the cross correlation function at $\pm \tau$. In addition, we also
observe correlation, although smaller, at  $\pm 3 \tau$,  $\pm 5
\tau$, etc.  On the other hand, when computing the cross
correlation function between {\em different} longitudinal
laser-cavity modes, we observe almost no correlation as can be
seen in Fig.~3, panel d) for the mode located at $\sim$ -200 GHz
in one laser and the one located at $\sim$ -70 GHz in the other
laser. This fact indicates that the synchronization takes place
only between the same longitudinal modes of the two lasers while
the correlation between different longitudinal modes is rather
weak. The fact that the same longitudinal mode of the different
lasers synchronize was also observed in a system of two
unidirectionally coupled semiconductor lasers \cite{White99}.

As in the experiments and previous numerical simulations
\cite{tilmann},\cite{josep} we also observe synchronization at
sub-nanosecond time scale. However, the quality of the
synchronization depends on the bandwidth of the detector. In
Figure~4 we plot the correlation coefficient, or the value of the
cross correlation function calculated at a time $\tau$, vs. the
bandwidth of the detector for both the main mode (dashed line) and
the total intensity (solid line). The synchronization is better
for the individual longitudinal modes than for the total intensity
and it extends almost over the whole range of detection without
loosing its quality. The partial loss of synchronization in the
slower-detector regime is due to the fact that the actual wave
forms emitted by the lasers consist of trains of rather short
pulses that are blurred when the detector response time is longer
than the typical pulse duration. On the other hand, the
synchronization gets worse for very fast detector as well. This is
because of the lack of synchronization between {\em different}
laser-cavity modes and by interference effects between them. As
can be noted in the figure, the detection bandwidth for the
isolated longitudinal mode is restricted to frequencies up to 10
GHz due to of the previous optical filtering process. In any case,
it is important to remark that a a high degree of synchronization
is obtained for a wide detection bandwidth.

Finally, we considered a similar situation that we have already
discussed but the lasers are now two DFB lasers. They are placed
at the same distance and pumped close to threshold. The observed
behavior of the output power is qualitatively similar to the one
showed in Figure~1 a). However, we observe in the optical spectra
that the lasers operate mainly in one longitudinal mode and only
one side mode carries a small fraction of power. As in the FP
case, the spectra of both lasers are very similar to each other,
indicating a high degree of synchronization. After filtering the
longitudinal modes we compute again the cross correlation
function. In Figure~5 we plot this function for the two modes and
for the total power. As expected, there are only small differences
between the cross correlation of the total power and the one of
the main longitudinal mode. But it can be also seen that the side
modes synchronize to its counterpart at the same time shift $\pm
\tau$, $\pm 3\tau$, etc. as the total power or the main mode power
do. This again indicates that the synchronization takes place at
the same mode of the different lasers.

The important difference from the Fabry-Perot system case is that
with the DFB lasers we can identify the laser mode that is
responsible for the LFF behavior. Moreover, we can directly check
if the other mode, the suppressed one, plays any role in the
destabilization process. In Fig.~6 we plot, for comparison, the
time-dependent modal powers during two consecutive dropouts. It is
important to point out that for the time traces we have a time
resolution of $\sim$ 0.1 ps. It can be clearly seen that the side
mode typically exhibits measurable power only after a power
dropout of the main mode develops. After the main mode recovers,
the side-mode power steadily decreases until the next dropout,
increasing the side-mode suppression ratio to several orders of
magnitude. That is a strong indication that the side mode is
actually not important for the LFF behavior and does not play any
role in triggering the power dropouts. By the same token, it is
also a strong indication that the single-mode models actually do
capture the essential physics of the phenomenon.

In conclusion we have carried out a study of the dynamics of two
distant, mutually coupled semiconductor lasers. To describe the
lasers we have used a laser simulator with full spatial and
temporal resolution that captures correctly the dynamics of both,
Fabry-Perot and DFB lasers  and includes a realistic model for the
active medium. We have observed synchronization between the two
output powers when one of the series is shifted with respect to
the other by a time $\tau$ corresponding to the external cavity
length. By filtering individual laser-cavity modes we have
observed that this synchronization takes place between the the
same individual longitudinal mode of the two lasers. On the other
hand, the degree of synchronization between {\em different}
laser-cavity modes turns out to be much smaller. As a consequence,
the quality of the synchronization is better for the individual
longitudinal modes than for the total power. We have also studied
coupled DFB lasers to compare a multi-mode regime with an
essentially single-mode situation. Our findings indicate that the
dynamics responsible for the LFF behavior and the output power
synchronization takes place within a single laser-cavity mode.
Moreover, we have also observed that the suppressed mode does not
play any role in triggering the LFF.

This work was funded by Spanish MCyT under projects
TIC99-0645-C05-02 and BFM2000-1108, by DGES under project
PB97-0141-C02-01 and also by AFOSR grant no. F4962-00-1-0002 and
AFOSR DURIP grant no. F4962-00-1-0190. M.K. was partly supported
by the GASR grant VEGA 2/7174/20.

\pagebreak

\section*{Figure Captions}
\vspace{1cm}

 \noindent Figure 1- Output intensity (with a
detection bandwidth of 0.5 GHz) of the two coupled Fabry-Perot
lasers (a), and their time-averaged optical spectra (b).
\vspace{1cm}

\noindent Figure 2- Synchronization of the output powers in the
most-intense laser-cavity mode for both lasers. One of the outputs
(showed in dashed line) is delayed by the external cavity trip
time $\tau$.
\vspace{1cm}

\noindent Figure 3- Cross-correlation function of the output
powers of the Fabry-Perot laser twins in the a) total output, b)
most intense laser-cavity mode, and c) one of the weak side modes.
Panel d) shows the cross-correlation between different laser
cavity modes, with no significant synchronization present.
\vspace{1cm}

\noindent Figure 4- Correlation coefficient as a function of the
detector bandwidth. The full curve (and star symbols) corresponds
to the total output power, while the dashed line (and diamond
symbols), that lasts until 10 GHz, shows the result for a single,
filtered laser-cavity mode.
\vspace{1cm}

\noindent Figure 5- Cross-correlation function of the output
powers of the two coupled DFB lasers. Panels a) and b) show the
total output powers and the main mode correlations, respectively.
Panel c) shows the cross-correlation functions between the
side-modes of the two lasers.
\vspace{1cm}

\noindent Figure 6- Output power of the dominant mode (upper
curve) and of the side mode (lower curve) of one of the DFB laser.
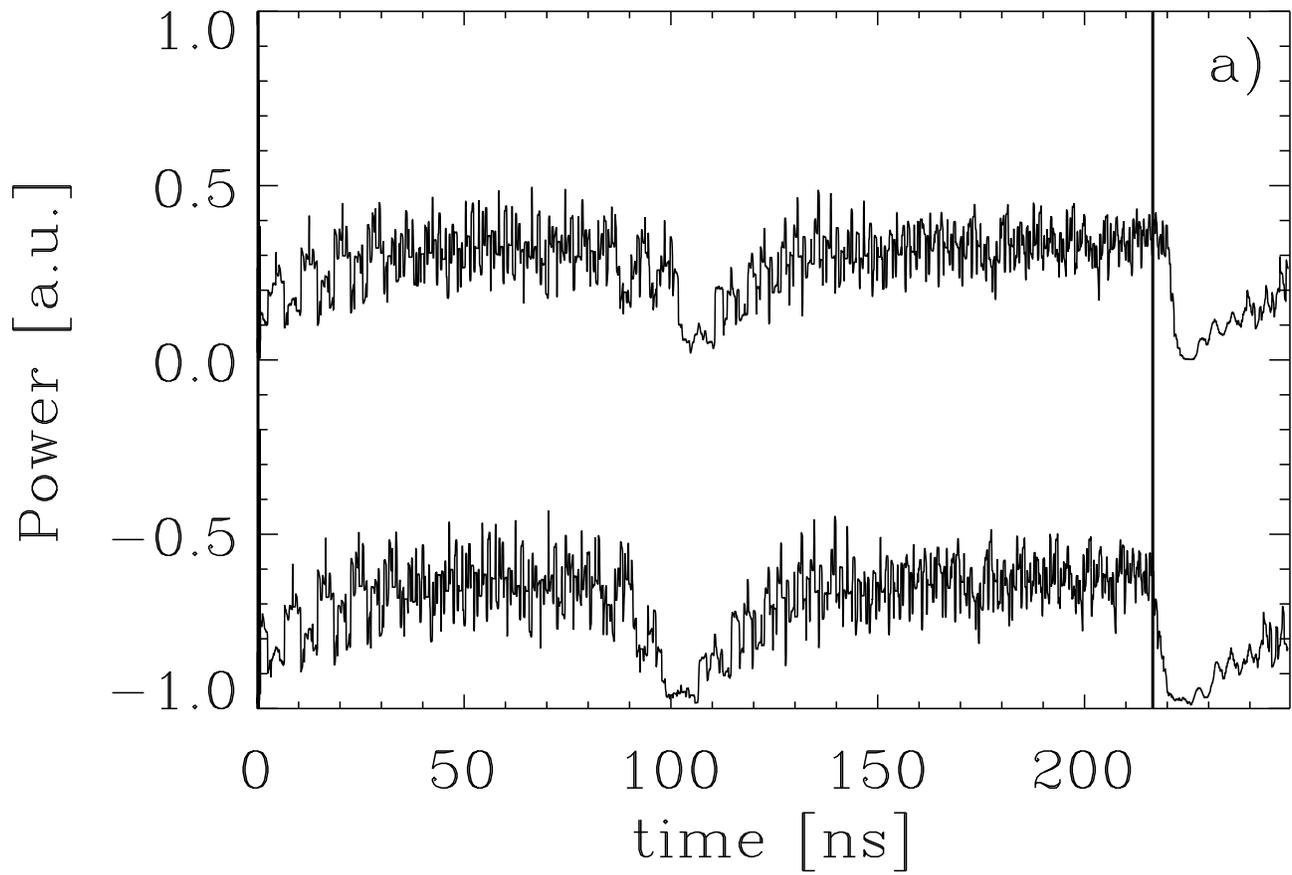
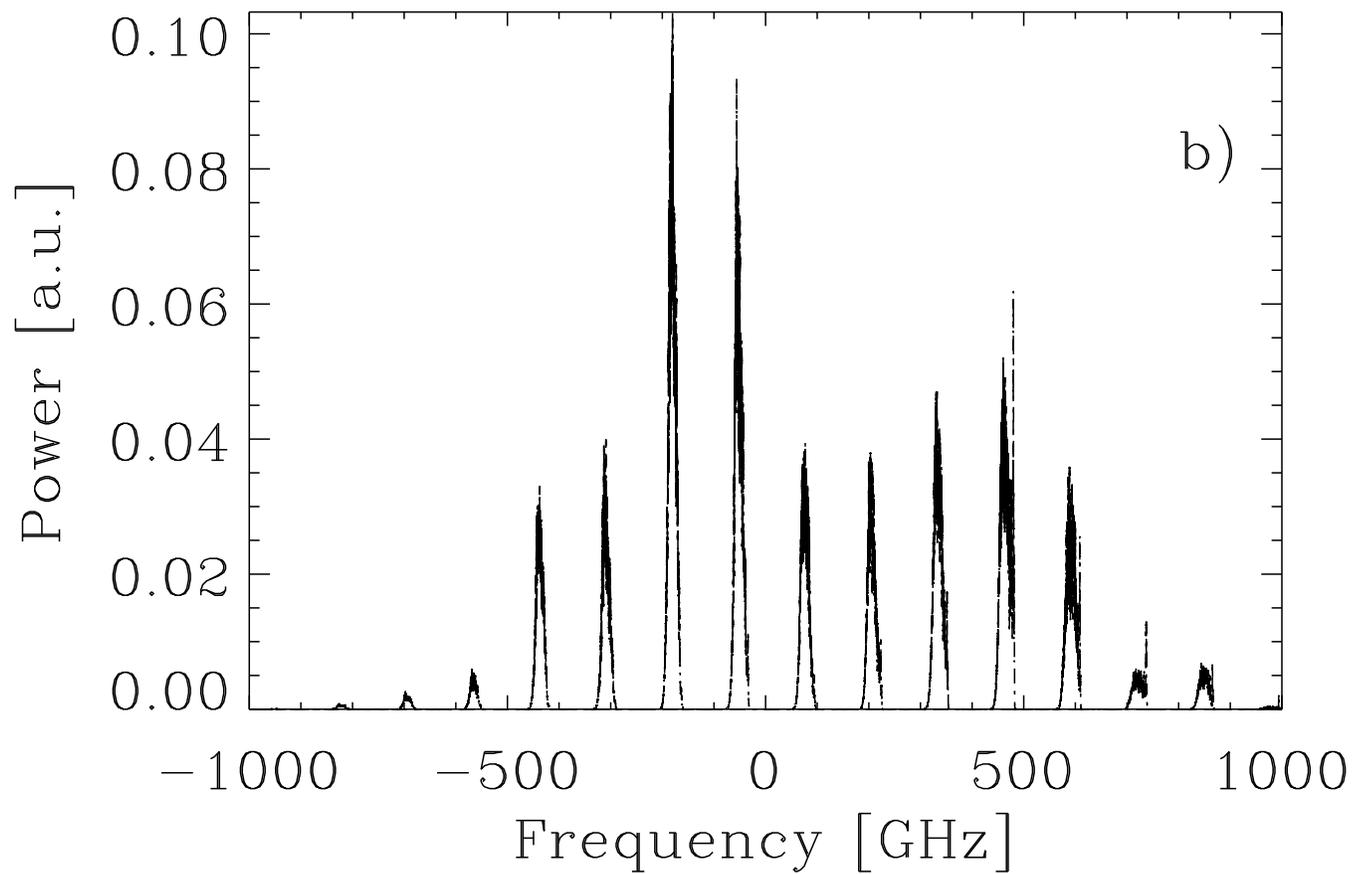
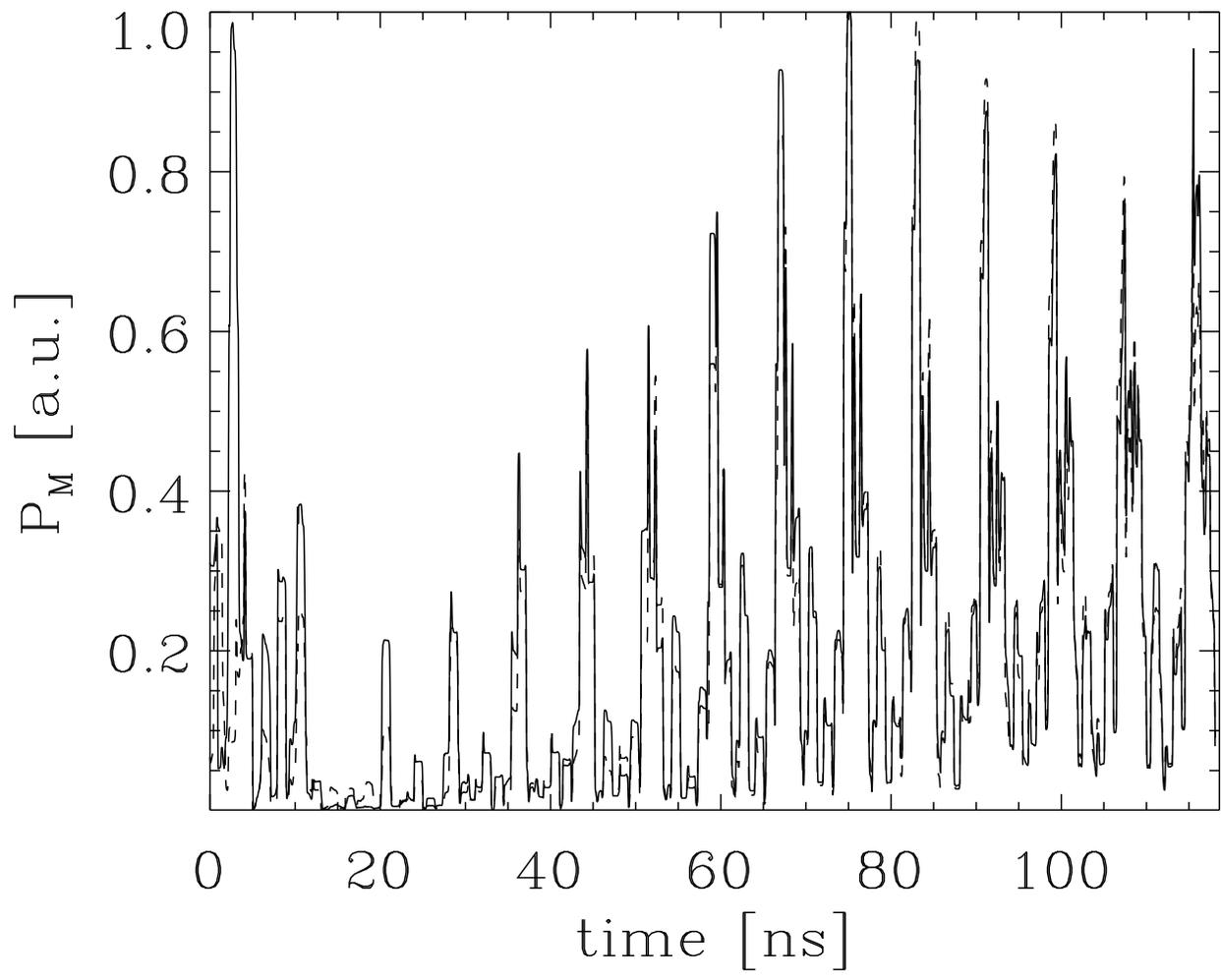
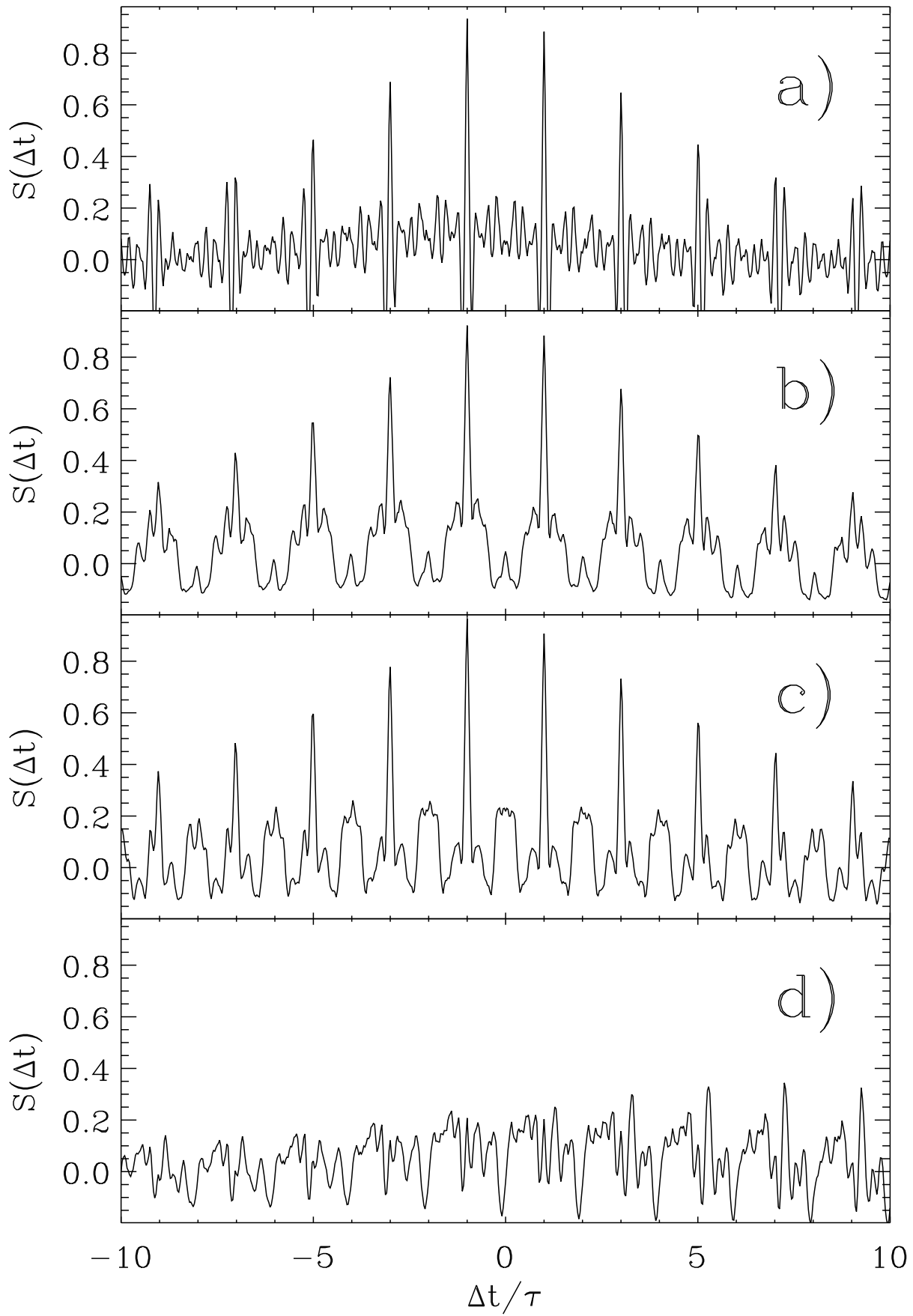
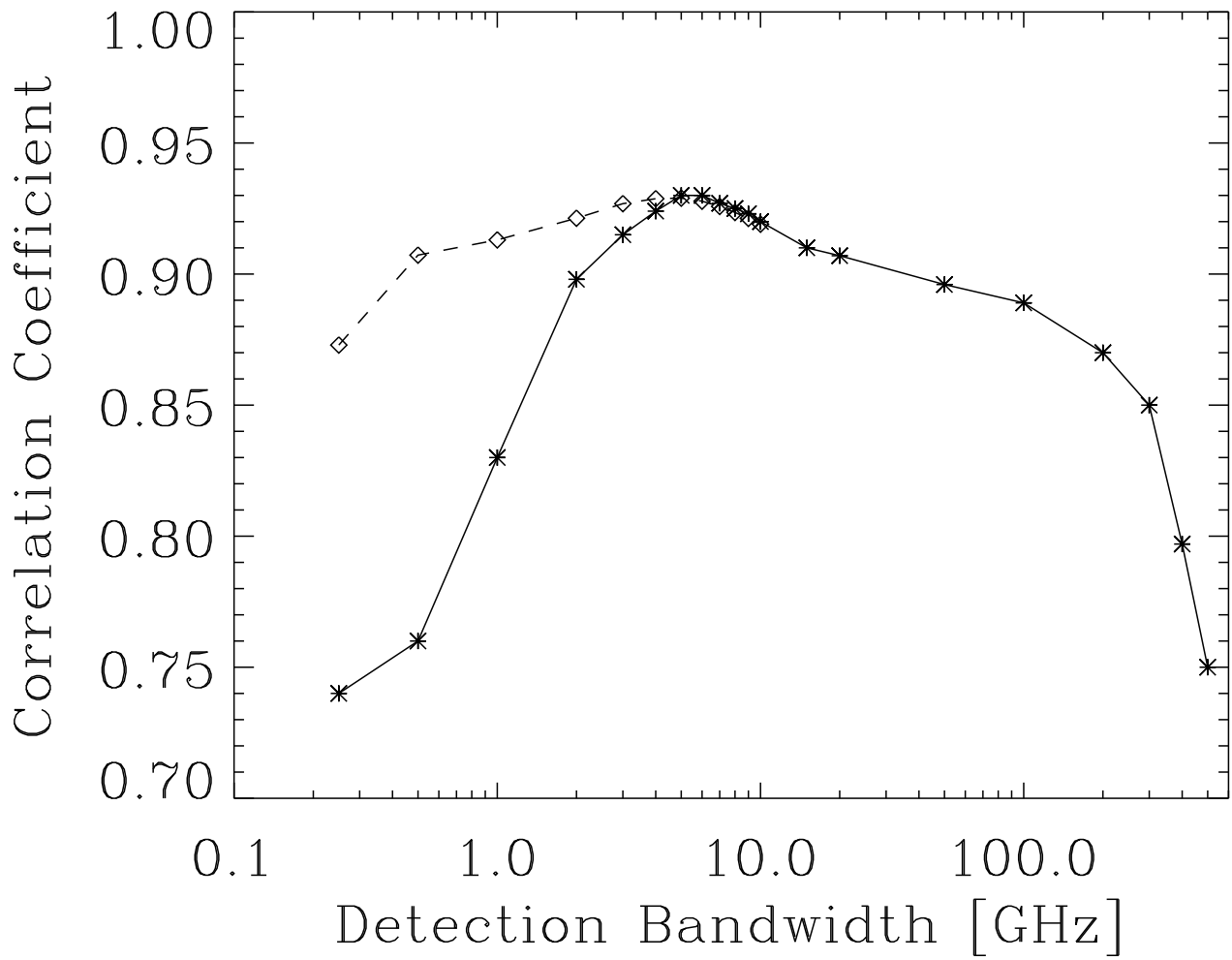
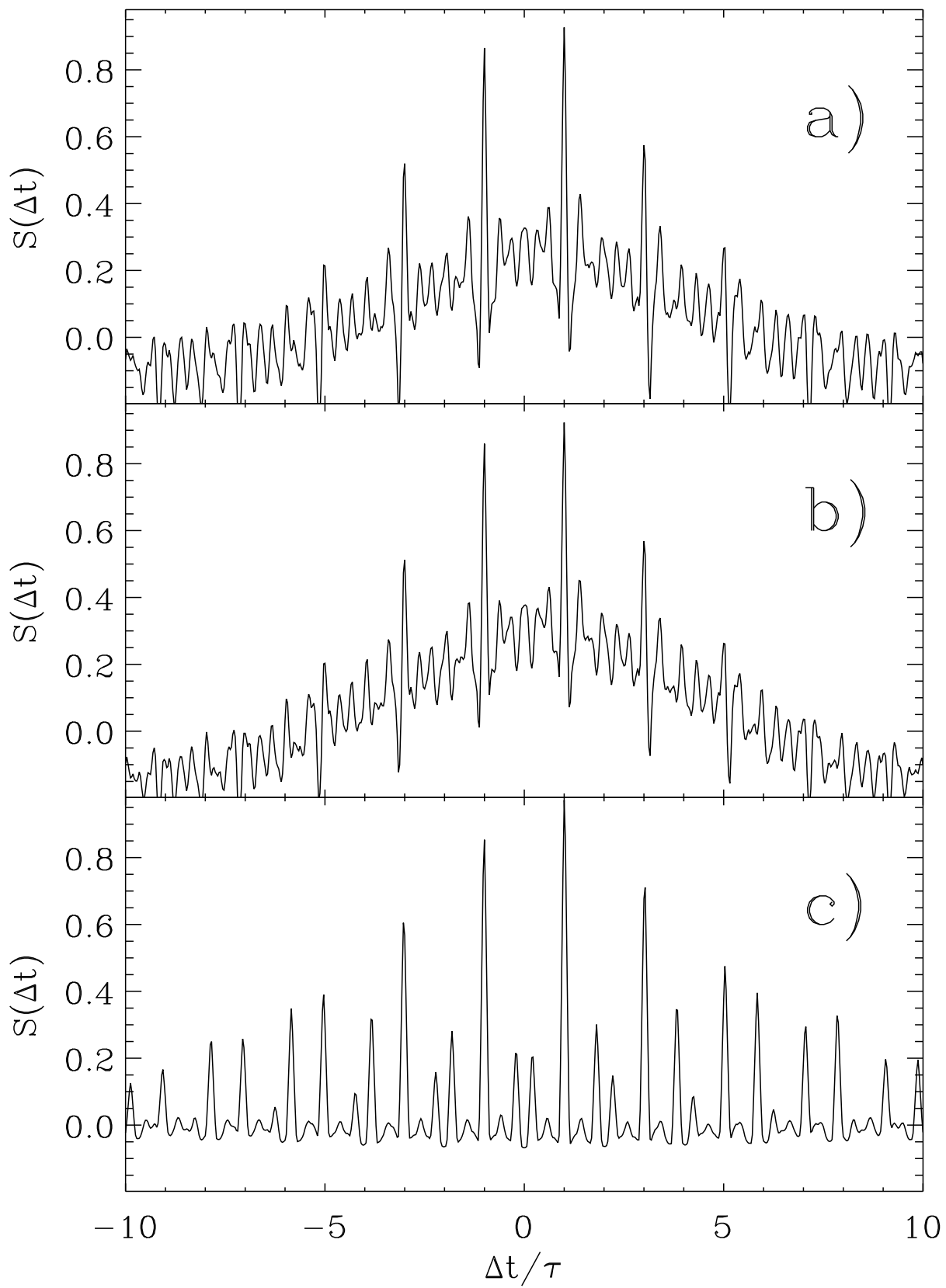
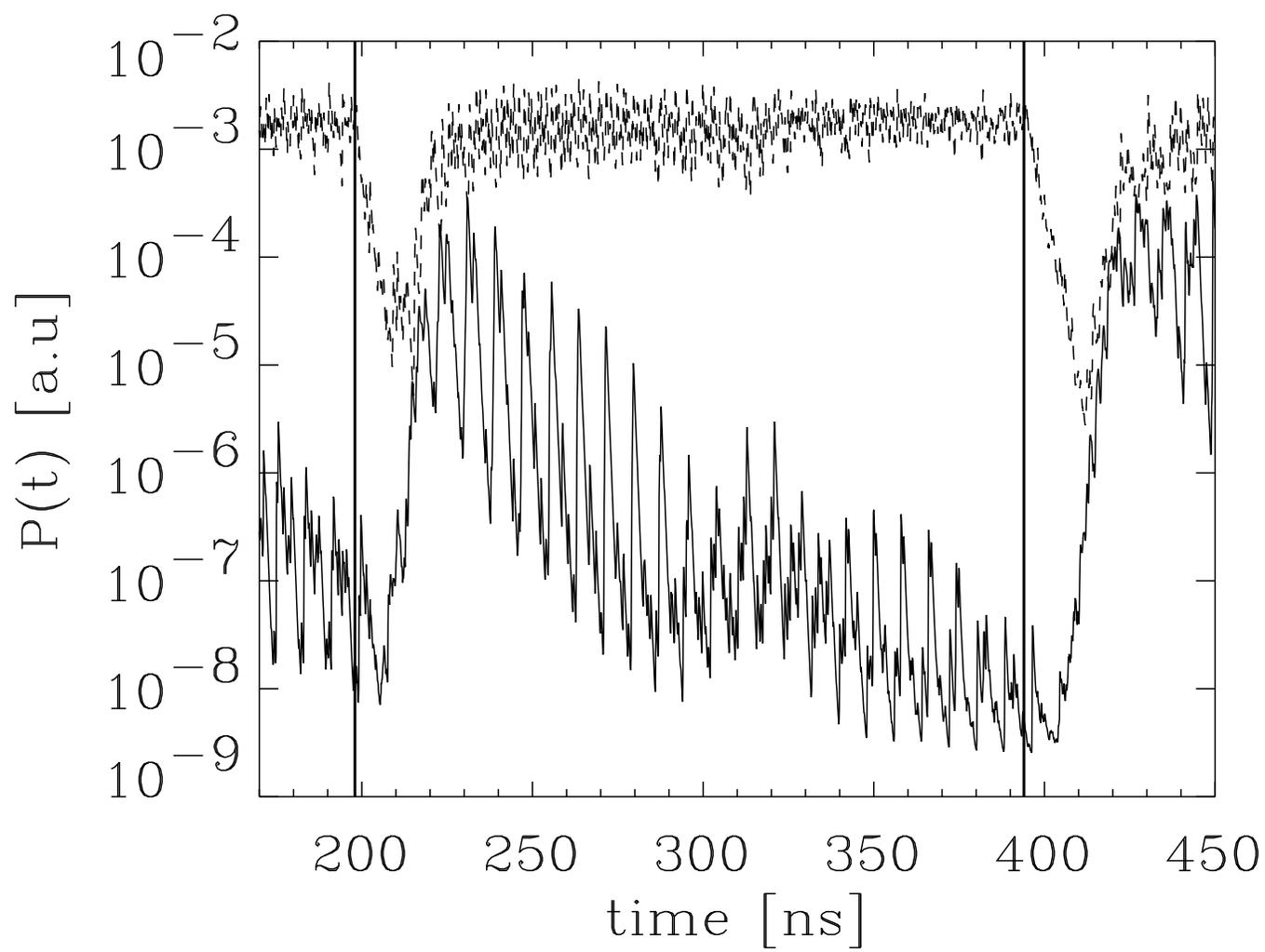
\end{document}